\documentclass[conference]{IEEEtran}
\IEEEoverridecommandlockouts
\usepackage{amsmath,amssymb,amsfonts}
\usepackage[top=25mm, bottom=25mm, left=20mm, right=20mm]{geometry}
\usepackage{algorithmic}
\usepackage{graphicx}
\usepackage{textcomp}
\usepackage{xcolor}
\def\BibTeX{{\rm B\kern-.05em{\sc i\kern-.025em b}\kern-.08em
    T\kern-.1667em\lower.7ex\hbox{E}\kern-.125emX}}
\usepackage{caption}
\usepackage{subcaption}
\usepackage{chemformula}
\usepackage{multirow}
\usepackage[labelsep=period]{caption}
\usepackage[style=ieee,backend=bibtex,bibencoding=ascii,sorting=none,maxbibnames=99]{biblatex}
\bibliography{bibliography}

\begin{document}

\title{Fair Reinforcement Learning Algorithm for PV Active Control in LV Distribution Networks}

 \author{\IEEEauthorblockN{Maurizio Vassallo\IEEEauthorrefmark{1},
 Amina Benzerga\IEEEauthorrefmark{1},
 Alireza Bahmanyar\IEEEauthorrefmark{1},
 Damien Ernst\IEEEauthorrefmark{1}\IEEEauthorrefmark{2}
}
 \IEEEauthorblockA{\IEEEauthorrefmark{1} Department of Electrical Engineering and Computer Science,
 Liège, Belgium
 }
 \IEEEauthorblockA{\IEEEauthorrefmark{2} LTCI, Télécom Paris, 
 Institut Polytechnique de Paris,
  France}
 \IEEEauthorblockA{\{mvassallo, abenzerga, abahmanyar, dernst\}@uliege.be}
 % \thanks{This research is supported by the public service of Wallonia within the framework of the Silver project.}
 }

\maketitle

\begin{abstract}
The increasing adoption of distributed energy resources, particularly photovoltaic (PV) panels, has presented new and complex challenges for power network control. With the significant energy production from PV panels, voltage issues in the network have become a problem. Currently, PV smart inverters (SIs) are used to mitigate voltage problems by controlling their active power generation and reactive power injection or absorption. However, reducing the active power output of PV panels can be perceived as unfair to some customers, discouraging future installations.
To solve this issue, in this paper, a reinforcement learning technique is proposed to address voltage issues in a distribution network, while considering fairness in active power curtailment among customers. The feasibility of the proposed approach is explored through experiments, demonstrating its ability to effectively control voltage in a fair and efficient manner.
\end{abstract}


\section{Introduction} \label{introduction}
% Context: The background of your topic
    %% current situation
The increase in global energy consumption, the unstable cost of electricity, and the goal of reducing \ch{CO2} emissions are among the most important reasons for the vast integration of distributed energy resources (DERs) throughout distribution networks.\\
% Scope: The topic you’ll be covering
    %% introduction of DER (advantages)
% These distributed devices are able to generate electricity from natural resources with low carbon emissions. Their main advantages, other than low emission electricity generation, is to reduce energy losses, since they are usually close to the consumption point, and helping decreasing the stress on the grids, especially during high-peak production.
    %% what problem they cause (disavantages)
However, due to the decentralised production of electricity, grid management has become more complex, demanding more sophisticated control methods to guarantee grid reliability.
This presents a challenge for a distribution system operator (DSO), who needs to ensure that power networks operate within their operational limits. One of the outcomes of the widespread integration of DERs is the occurrence of reverse power flow that occurs when the energy generated exceeds the energy consumed. 
This reverse power flow can result in certain voltage-related problems.
Voltage management in electricity distribution networks has been widely studied in the literature. In \cite{OLTC}, the authors use a fuzzy logic algorithm to control the tap position of the transformers. Although this solution may be effective for traditional distribution networks, its implementation in DER-dominated networks is impractical due to the unpredictable and fluctuating nature of the DERs' output. This would require frequent tap adjustments, resulting in high maintenance costs.
% For this reason, some other works have tried to find configurations and techniques to minimise the number of tap changes \cite{OLTCmainte, NCT}.
Another possible solution would be to use the flexibility of DERs and controllable loads. For example, using an optimal power flow (OPF) technique, it is possible to optimise the operation of power systems by adjusting the generation and load levels to minimise the overall cost of operation while meeting various operational constraints. This involves solving a large-scale optimisation problem that considers multiple nonlinear operational constraints and objectives, making it computationally expensive \cite{OPF}.\\
% Some works deal with improving the performance of the OPF calculation \cite{OPVIRCP}.\\
% Smart inverters (SIs) have gained popularity in recent years due to the increasing penetration of DERs. SIs can be used to regulate the voltage of the network, controlling their active and reactive power outputs. Several solutions have been proposed to control the voltage with SIs. When dealing with SIs, it is possible to control either the active power, reactive power, or both. \\
%Active
In \cite{app11177900}, the authors propose an active power curtailment technique with different types of droop-based methods and a dynamic programming method to minimise power curtailment. Authors in \cite{HOWLADER2020114000} discuss the potential of smart inverters (SIs) to mitigate voltage and frequency deviations using active power curtailment, volt-watt control, and frequency-watt control. The active power curtailment control can be one of the most effective controls for voltage issues. However, this technique reduces the energy output of the DERs to the grid, resulting in economic drawbacks for customers. 
%Reactive
For this reason, some works have preferred to control the reactive power instead.
Paper \cite{IVRUDN} presents a reactive power-based control strategy for single-phase PV inverters to improve voltage unbalance and voltage regulation in low-voltage distribution networks. Similarly, in \cite{DVVCPVI}, authors propose a voltage control loop that can absorb or supply reactive power to maintain voltage within acceptable bounds. However, only reactive power control is not always enough to solve the voltage problems. 
%Both
Consequently, some studies have focused on addressing both active and reactive control to mitigate voltage-related issues. The method proposed in \cite{AMLVNMO} provides an effective voltage regulation while decreasing power losses and maximising the revenue of the customers. The authors in \cite{DVRAPC} propose a new technique for mitigating over-voltage violations in distribution networks, using short-term PV power generation forecasts to adjust active and reactive power injection from the PV inverters.\\
% With the improvements in artificial intelligence. In particular, thanks to the results of reinforcement learning (RL) in other fields \textcolor{red}{cite fields}, more and more solution are applied using RL techniques have been applied on power networks problems. Some solutions have been applied with DQN, DDQN, DDPG, SAC \textcolor{red}{cite papers for each algo}.\\
All the solutions discussed so far have tried to solve the voltage problems without dealing with fairness during curtailment. Indeed, the curtailment of DERs is dependent on their location, and customers might be curtailed unequally and unfairly \cite{TONKOSKI20113566}. In \cite{CAPCGC}, the authors propose to share the required curtailment equally among the customers.\\
% Their results showed that sharing the power curtailment among all customers comes at the cost of an overall higher amount of power curtailed.\\
Many definitions of fairness have been introduced. Authors in \cite{CFCGDVCMPI} introduce a definition of fairness and apply it in \cite{PIBFPQC} where they propose a complex analytical approach for achieving fair active power curtailment and reactive power control in low-voltage distribution networks. Simpler solutions are proposed using reinforcement learning (RL) techniques. In \cite{MFAFSPCURF}, the authors apply a soft actor-critic (SAC) based RL algorithm to address the voltage issues using three definitions for fairness. The authors consider only active power curtailment, that could lead to suboptimal solutions. \\

%Paper goal
The objective of this study is to explore the implementation of a deep RL technique for managing the active and reactive power outputs of SIs in an LV distribution network. The proposed approach aims to ensure fairness among the customers while complying with the voltage regulation and minimising active power curtailment. The main benefit of this RL-based approach lies in its capacity to deliver an optimal and fair solution with a limited amount of network data requirements.\\

%Paper strucuture
The rest of the paper is organised as follows. Section \ref{problemstatment} defines the problem. Section \ref{bk} presents some background information about fairness, SI, and RL, respectively. Section \ref{implementation} presents the proposed method. The results are reported in Section \ref{results} and Section \ref{conclusion} concludes the work and discusses some possible future works.

% \section{Literature Review} \label{litrev}
% \input{sections/LiteratureReview.tex}

\section{Problem Statement} \label{problemstatment}
% One of the tasks of DSOs in maintaining the power network within its operational limits. In particular, DSOs must accurately assess whether the network is in a critical state at a given moment and take appropriate measures to mitigate voltage issues. These measures may include applying active power curtailment of generators and controlling their reactive power.\\

% \noindent For this problem, we assume that the DSO knows the following information:
% \begin{itemize}
%     \item The network topology: the number of buses, loads and generators, the lines' length, the distance between the connected buses, and the distance between each load and generator from the bus they are connected to. Moreover, the impedance of the lines is known.
    
%     \item The active and reactive power of loads and PV at each time step. 
% \end{itemize}

% \noindent This data is used to calculate the power flow of the network to obtain more information; like the voltage magnitude at each bus and the lines' loss.\\

A power network can be represented as a directed graph, $\mathcal{G} = (\mathcal{N},\mathcal{E})$, with a set of nodes, $\mathcal{N}$, also called buses. Directed edges, $\mathcal{E}$, link two buses together, also called lines. A line $e_{i,j} \in \mathcal{E}$ connects node $i$ to node $j$, with $i,j \in \mathcal{N}$.
Several devices, which may inject or withdraw power from the grid, might be connected to each bus. The sets of all devices, either loads $\mathcal{L}$ or generators $\mathcal{DG}$, are assumed to be known.  \\
% https://arxiv.org/ftp/arxiv/papers/2102/2102.05657.pdf
The DSO takes into account the behaviour of the network over a set of discrete time steps, $t$, over a studied time horizon, $T$.\\
We denote the active and reactive power of load $i$ at time step $t$ as $\mathcal{L}^p_{i,t}$ and $\mathcal{L}^q_{i,t}$, with $i \in \mathcal{L}$; the active and reactive power of the PVs as $\mathcal{DG}^p_{i,t}$ and $\mathcal{DG}^q_{i,t}$ with $i \in \mathcal{DG}$; the voltage of the node $i$ at time step $t$ as $V_{i,t}$, with $i \in \mathcal{N}$.\\
% Without lack of generality, in this paper, the network topology is considered to be static, so that no reconfigurations are applied on the network during the whole time series' horizon.\\

% The goal of the DSO is to predict whether the system, at some given future time step, $t \in T$, will be in critical condition and take some actions to prevent it to happen.\\
The network situation, in a given time step $t \in T$, is considered critical if the voltage of at least one of the buses is outwith the boundaries: $V_{i,t} < v^{\text{min}}$ (under-voltage) or $V_{i,t} > v^{\text{max}}$ (over-voltage) for $\forall i \in \mathcal{N}$, where $v^{\text{max}}$ and $v^{\text{min}}$ are the maximum and the minimum acceptable voltage, respectively.

\section{Contextual background} \label{bk}
\subsection{Fairness} \label{fr}
Active power curtailment is not desirable from an economic and  environmental point of view because it implies that some energy cannot be produced, and that almost the same amount of the curtailed DER production needs to be replaced by another energy source. However, sometimes it is needed to curtail the energy production to keep the distribution network functional.\\
The amount of energy curtailment needed depends on many factors, such as the location of the DER in the feeder, the amount of the energy consumption and the amount of energy generation. 
Therefore, customers may not be curtailed equally, which might feel unfair and could be discouraging for new customers.\\

In this paper, we define fairness in active power curtailment as the equality in the amount of energy that each customer is being curtailed in case of a voltage issue. Fairness is evaluated using the Gini
index, which it is used to test the degree of inequality during curtailment. A low Gini index indicates a more equal distribution of curtailment, while a high Gini index indicates a more unequal distribution. In particular, a Gini index of $0$ reflects perfect equality, where all customers are curtailed the same, while a Gini index of $1$ reflects maximal inequality among customers' curtailment \cite{enwiki:1148472029}.

\subsection{Smart inverters} \label{si}
Smart inverters for PV systems are electronic devices that convert the direct current (DC) produced by solar panels into alternating current (AC). Unlike traditional inverters, smart inverters incorporate advanced features and communication capabilities that allow them to actively manage and control the flow of electricity from the PV system into the grid. This enables them to regulate the voltage and frequency of the AC power, which helps maintain reliability and stability.
The control is performed adjusting their active and reactive power outputs.\\
\begin{figure}[h!]
    \centering
    \includegraphics[width=0.33\textwidth]{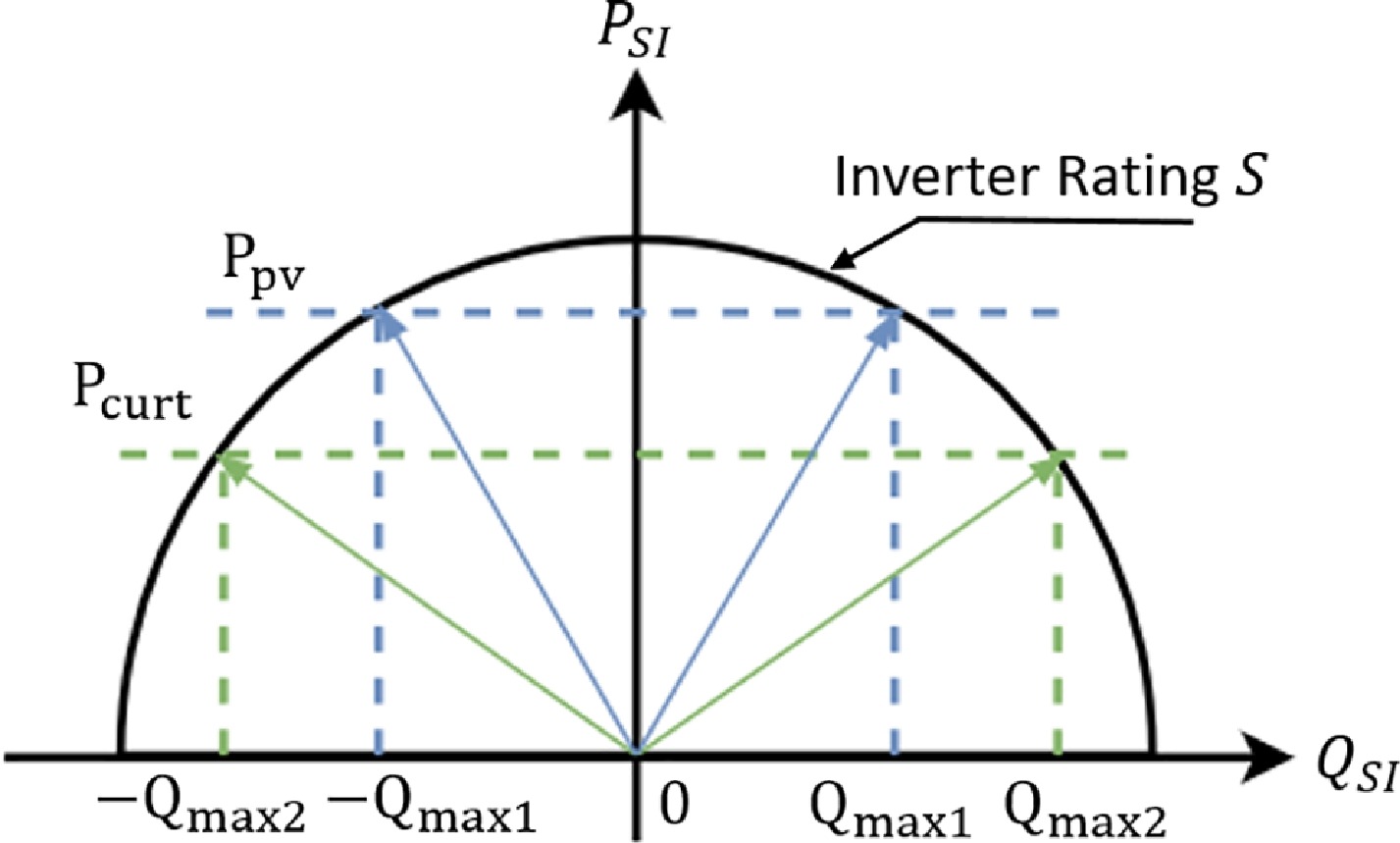}
    \caption{SI capability curve \cite{lr_csi}}
    \label{fig:si}
\end{figure}
\noindent As shown in Fig. \ref{fig:si}, the maximum output of a SI is constrained by its rated output $S$. The active power, $P$, is mainly dependent on the weather conditions. The active power, $P$, and reactive power, $Q$, can be controlled to satisfy the formula $P^2 + Q^2 \leq S^2$. Therefore, in situations where critical voltage issues necessitate a certain amount of reactive power, $\pm Q_{max2}$, that exceeds the current available quantity, $\pm Q_{max1}$, and the active power output is already high, $P_{pv}$, curtailing the active power, $P_{curt}$, can create additional capacity for reactive power. The active power curtailment, $P_{curt}$, is constrained by $0 \le P_{curt} \le P_{pv}$.\\
In practice, however, the SIs have different capability curves and operational constraints, as documented in the literature \cite{Nouri_2020}.

\subsection{Reinforcement learning} \label{rl}
\begin{figure}[h!]
    \centering
    \includegraphics[width=0.42\textwidth]{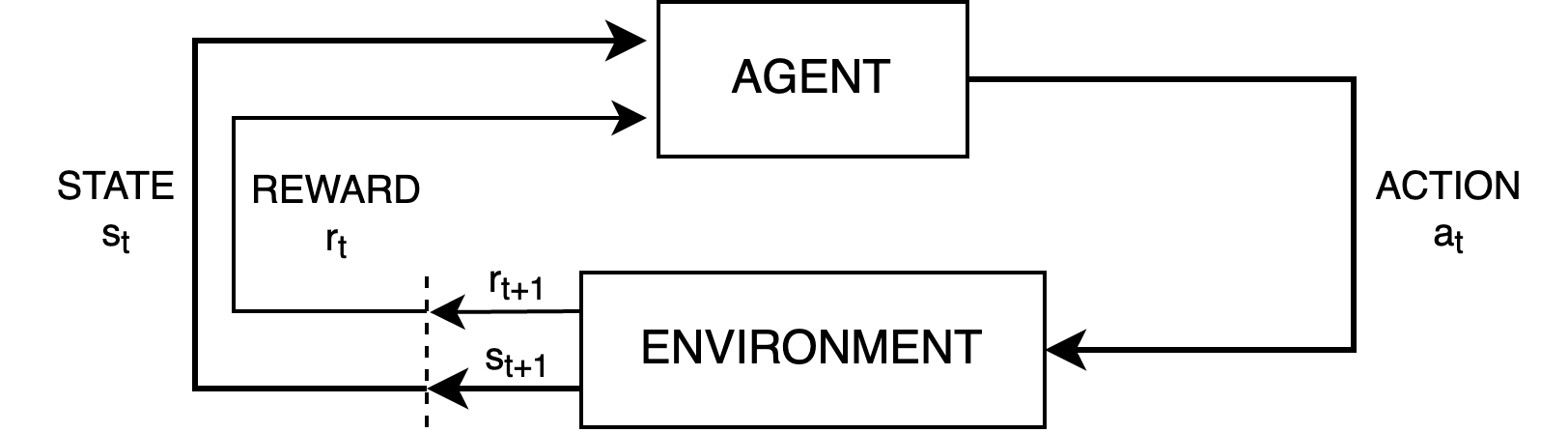}
    \caption{Action-reward feedback loop of a generic RL framework \cite{sutton2018reinforcement}}
    \label{fig:rl}
\end{figure}

In an RL framework, an agent interacts with the environment over a series of discrete time steps represented by $t \in T$, where $T$ denotes the set of all the time periods under consideration. As shown in Fig. \ref{fig:rl}, at each time step $t$, the agent receives some information from the environment, the state $s_t$, takes an action $a_t$, and receives a numerical feedback from the environment, the reward $r_t$. With the action chosen, the agent modifies the environment, ending up in a new state $s_{t+1}$.\\
The agent learns by repeatedly executing a loop of state-action-reward-new$\,$state. Through this process, the agent becomes better at selecting actions that lead to positive rewards and avoiding actions that result in negative rewards.\\
% This loop can be seen as a Markov decision process (MDP), indeed, a reinforcement learning problem is modelled as a MDP and described by the tuple $(\mathbb{S}, \mathbb{A}, \mathbb{P}, \gamma, \mathbb{R})$, where:
% \begin{itemize}
%     \item $\mathbb{S}$ is the state space
% \end{itemize}

\noindent The goal of the agent is to maximise the expected return:
\begin{equation}
R_t = \lim_{T \to \infty} \mathbb{E}_{\pi} \big[ \sum_{i=t}^{T} \gamma^{i-t} r_i \big]
\end{equation}
where $\pi$ is the policy used to choose the action $a_t$ given the state $s_t$ and $\gamma$ is the discount factor, $0\leq\gamma\leq1$, that it is used to express how the future rewards should be considered when making decisions in the present.\\
The action-value function, also known as Q-function, is defined as the expected return when the agent in state $s_t$, decides to take the action $a_t$, following the policy $\pi$:
\begin{equation}
Q_{\pi}(s_t,a_t) = \lim_{T \to \infty} \mathbb{E} \big[ \sum_{i=t}^{T} \gamma^{i-t} r_t|s_0=s_t,a_0=a_t\big]
\end{equation}
The general RL solution makes use of the recursive relationship known as the Bellman equation:
\begin{equation}\label{eq:belman}
Q_{\pi}(s_t,a_t) = \lim_{T \to \infty} \mathbb{E}
\big[
    r_t + 
    \gamma \mathbb{E} \big[ Q_{\pi}(s_{t+1},a_{t+1}) \big]
\big]
\end{equation}
This recursive relationship allows the agent to estimate the value of a state by considering the values of its future states and their associated rewards.\\
In standard Q-Learning, the action $a_{t+1}$ in Eq. \ref{eq:belman} is chosen with a greedy policy $a_{t+1} = \mu(s_{t+1}) = argmax_a Q(s_{t+1},a)$, which chooses the action that maximises the Q-value of the next state. This approach has been widely used in RL and has proven to be effective in a wide range of applications \cite{li2018deep}.\\

% \subsubsection{DDPG}
The greedy approach adopted by Q-Learning to optimise the policy $\pi$ renders it unsuitable for continuous action spaces, as the $argmax$ operation is a limiting factor.\\
The deterministic policy gradient (DPG) implementation tries to solve this problem. The idea behind DPG is to learn a deterministic function, $\mu(s_t)$, that can approximate the $argmax_a Q(s_t,a_t)$ operation. Generally, the $\mu(s_t)$ function is constructed as a deep neural network (Deep DPG) \cite{lillicrap2019continuous}.\\
In particular, DDPG is an actor-critic algorithm that uses two neural networks: an actor (characterised by weight parameters $\phi$) decides the action to perform in a given state, and a critic (characterised by weight parameters $\theta$) evaluates the action chosen by the actor.\\
\begin{figure}[h!]
    \centering
    \includegraphics[width=0.46\textwidth]{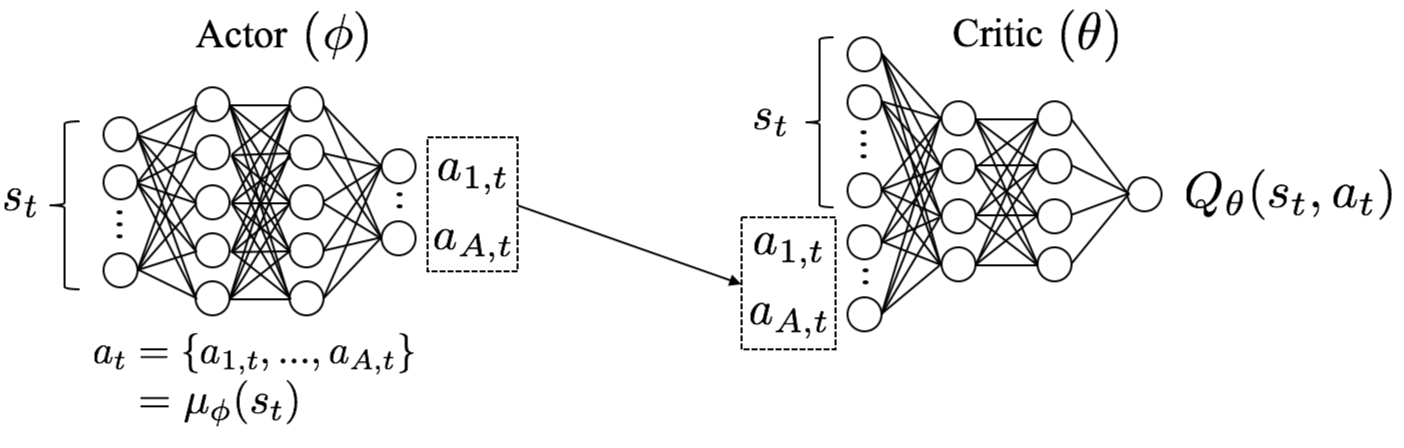}
    \caption{Actor-critic architecture in DDPG \cite{DRLBVCDMUDN}}
    \label{fig:ddpg}
\end{figure}
The critic is updated using the Bellman equation (Eq. \ref{eq:belman}). The actor is updated using the gradient of the expected return from a start distribution, J, with respect to the actor parameters. 
% Inspired by 'Coordination of PV Smart Inverters Using Deep Reinforcement Learning for Grid Voltage Regulation'
\[
\nabla_{\phi}J \approx
\mathbb{E} \big[ \nabla_{\phi} Q_{\theta}(s_t,a) | a=\mu_{\phi}(s_t) \big]
\]

Additionally, in order to enhance the performance of the model, various techniques are commonly applied in RL problems. These include memory buffers to store experiences, target networks to stabilise learning by using separate networks for estimation and actions selection, and batch normalization to improve the convergence of the model. 
% \textcolor{red}{Tabular version missing}

\section{Proposed method} \label{implementation}
\subsection{Distribution network}
Results are obtained on a distribution network with a single feeder. The network is a low-voltage distribution network, and it is served by a 0.4 MVA MV/LV transformer station. The network has 20 customers and each customer has a PV generator. Hence, for this work, the PV penetration rate is set to 100\%.\\
The time series are considered for one year with 15-minute resolutions, for a total of 35040 time steps. Different load profile time series are used for the customers, and similar generator profiles are used for the PV generators. The time series are taken from the SimBench dataset \cite{simbenchdataset}. 
\vspace{-1mm} %Without it it leaves a loooot of space *CLUELESS*
\subsection{RL elements}
\begin{enumerate}
    \item Environment:\\
The environment is the LV distribution grid as described in \ref{problemstatment}.

    \item State:\\
The state comprises the information accessible to the agent. In this case, the active and reactive power of loads, the active power of generators, and the voltage magnitude of each bus at time step $t$.   
\begin{equation}
    s_t = [\mathcal{L}^p_t, \mathcal{L}^q_t, \mathcal{DG}^p_t, V_t]
\end{equation}

The state space, $\mathcal{S}$, is given by all the possible combinations of $[\mathcal{L}^p_t, 
\mathcal{L}^q_t, 
\mathcal{DG}^p_t,
V_t]$. Thus, at any given instant, $s_t$ is a subset of $\mathbb{R}^n$, where $n$ is given by: $n = |\mathcal{L}| + |\mathcal{L}| + |\mathcal{DG}| + |\mathcal{N}|$.

    \item Action:\\
Similar to \cite{lr_csi}, the agent can choose the reactive power output of each SI, from $-S$ to $+S$.
As mentioned in Section \ref{si}, the action, $a_{i,t}$, can change the active and reactive power of each PV $i$, with $i \in \mathcal{DG}$.
In particular, the agent chooses an action $a_{i,t}$ to control the amount of reactive power, $Q_{i,t}$, of the SI $i$ at time step $t$. The reactive power chosen is given by $Q_{i,t} = S_i \cdot a_{i,t}$. If the value of $Q_{i,t}$ chosen by the agent satisfies the relationship $P^2_{i,t} + Q^2_{i,t} \leq S^2_i$, no curtailment is applied; otherwise some curtailment is applied and the new active power generation is given by: $\bar P_{i,t} = \sqrt{S^2_i - Q^2_{i,t}}$.\\
% The agent can not increase the value of the active power, $P_{i,t}$, so the next active power chosen by the agent, $\bar P_{i,t}$, is constrained by $0 \le \bar P_{i,t}<P_{i,t}$, where $P_{i,t}$ is the amount of power generated by the PV $i$ at time step $t$ and $\bar P_{i,t}$ is the new active power chose by the agent.\\
The action space, $\mathcal{A}$, is given by all the possible combinations of $[-1,1]^m$, where $m$ is given by: $m = |\mathcal{DG}|$.
    
    \item Reward:\\
The reward at time step $t$, $r_t$, consists of the sum of three terms:
    \begin{itemize}
    \item A reward, $r_v$, regarding voltage violation. The agent is punished if the voltage magnitude of the buses deviates from 1 pu.
    The idea is to punish the agent by a low value when the voltage bus is inside the range [$v^{min}$,$v^{max}$] and to punish it more when the voltage bus is farther away.
    This is ensured with a bowl-shaped voltage violation function, $f_v$, \cite{MARLAVCPDN}. For this paper, $v^{min}$ and $v^{max}$ are equal to 0.95 and 1.05, respectively. 
    % \cite{EN50160}.
    \begin{equation}
        r_v = - \sum_{i=0}^{\mathcal{N}} f_v(V_{i,t})
    \end{equation}

    % \begin{figure}[h!]
    %     \centering
    %     \includegraphics[scale=0.18]{figures/BVF.png}
    %     \caption{Voltage violation function $f_v$}
    %     \label{fig:TC}
    % \end{figure}

    \item A reward, $r_a$, limiting the magnitude of each action.
        \begin{equation}
            r_a = - \sum_{i=0}^{\mathcal{DG}} |a_{i,t}|
        \end{equation}
        % \[
        % r_a = - \sum_{i=0}^{\mathcal{DG} |\mathcal{DG}^q_{i,t}|
        % \]
    This is used to train the agent to choose actions that are as low as possible in magnitude to avoid too much curtailment and overuse of the reactive power.

    \item A reward for the fairness, $r_f$.
    \begin{equation}
        r_f = - \sum_{i=0}^{\mathcal{DG}} |\bar{c_t} - c_{i,t}|
    \end{equation}
    where $c_{i,t}$ is the active power curtailment of the PV $i$ decided by the agent when choosing the action $a_{i,t}$, and $\bar{c}_t$ is the mean value of the active power curtailment among the customers at time step $t$.\\
   The purpose of this reward is to encourage the agent to select actions that have relatively similar consequences on customers' curtailment so that if the agent decides to curtail one customer more than others, it receives a larger penalty.
    
    \end{itemize}
The total reward is given by the algebraic sum of all the previous terms multiplied by a scaling factor.
\begin{equation}
    r_t = \alpha \cdot r_v + \beta \cdot r_a + \omega \cdot r_f 
\end{equation}
The weights $\alpha$, $\beta$ and, $\omega$ are selected with careful consideration to balance the magnitude of each parameter, ensuring that the agent is able to optimise its performance while also exhibiting the desired behaviour. In this case study, the values for these parameters, considering the different scales of the single rewards, are chosen to be 1000, 5, and 25 respectively. These terms are chosen to have the following impacts on the reward function: $\alpha \cdot r_v \gg \beta \cdot r_a \simeq \omega \cdot r_f$.

\end{enumerate}

\noindent The agent's neural network architectures are composed of two hidden layers with 256 neurons each. The learning rate for the actor and critic network is set to 0.0001, the discount factor $\gamma$ is equal to 0.99 and the soft update parameter $\tau$ is equal to 0.005.
The batch size is 64 and the maximum number of experiences is 200000.
% During training, for the exploration the noise starts at 0.3, and it is reduced by 0.995 every 2000 time steps.

\section{Results}\label{results}
The results are presented for four different scenarios.
\begin{enumerate}
    \item Scenario \textit{a}. This is considered as the baseline: the situation where no control is applied and in the network are present some over and under-voltages that a DSO would like to solve.
    \item Scenario \textit{b}. The objective of the DSO is solely to address the voltage issues within the network. To reflect this goal in the reward function, only the $r_v$ term is taken into account.
    \item Scenario \textit{c}. The DSO aims to mitigate voltage problems while minimising active power curtailment. To reflect this objective in the reward function, both the $r_v$ and $r_a$ terms are taken into consideration.
    \item Scenario \textit{d}. The DSO wants to address the voltage problems while minimising active power curtailment and maintaining fairness in the distribution of the curtailment among customers. This is achieved by taking into account all three factors: voltage deviation, actions' magnitude, and fairness.
\end{enumerate}

\noindent In each scenario, the agent undergoes training across several episodes, with each episode covering 30\% of the total time steps $T$, corresponding to approximately the first 4 months of the year.
% The training is stopped as soon as the agent does not improve for four consecutive episodes.
After training, the agent is evaluated on the remaining 70\% of the time series, corresponding to approximately the remaining 8 months of the year.\\

\begin{figure}[h!]
    \centering
    \includegraphics[width=0.46\textwidth]{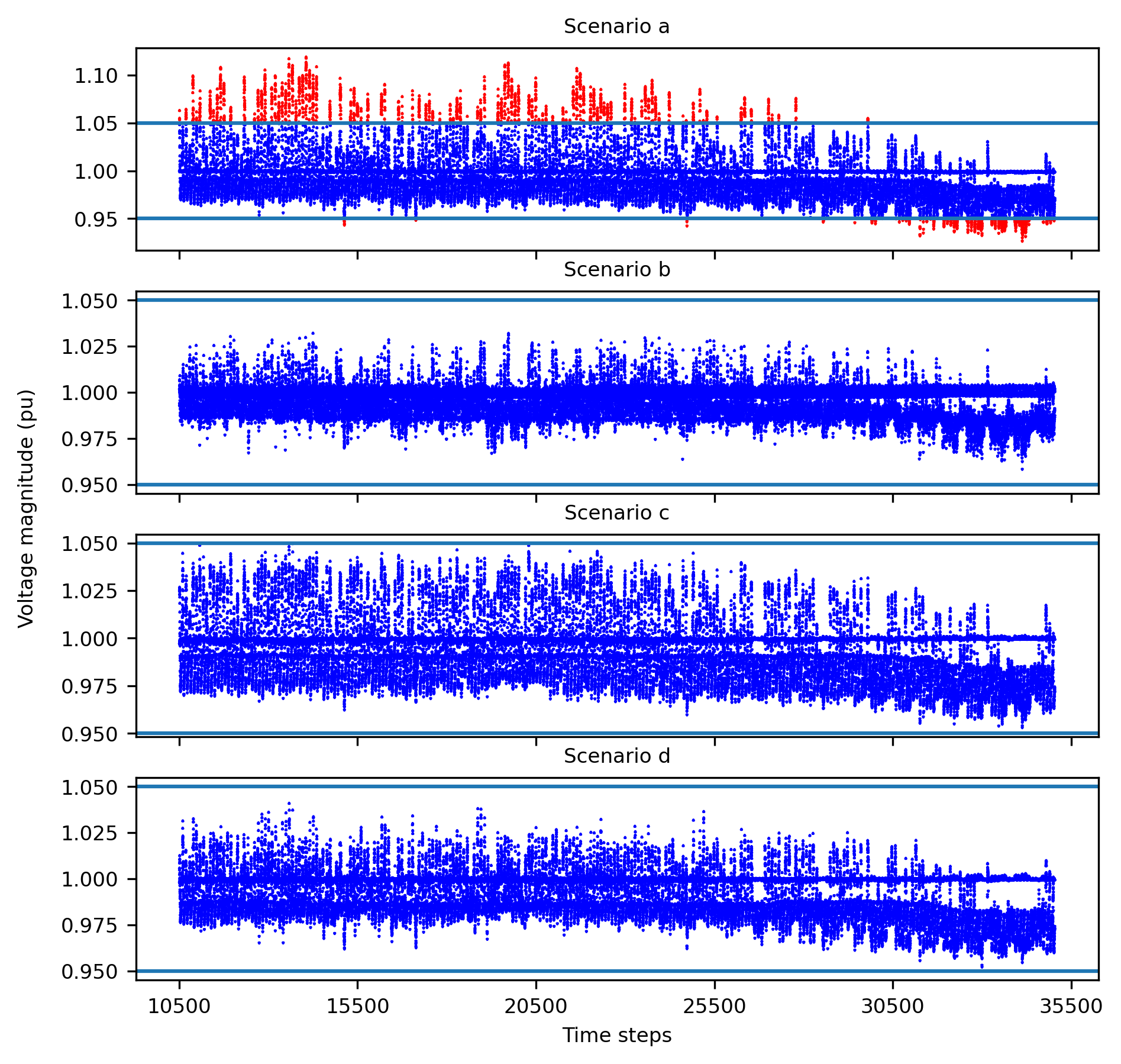}
    \caption{The buses' voltage situation for each scenario}
    \label{fig:Vsolv}
\end{figure}
Fig. \ref{fig:Vsolv} provides a visual representation of the voltage levels across the buses in each of the different scenarios. For each time step, two dots are drawn: representing the maximum voltage of all nodes recorded at that time step and a dot to represent the minimum. Moreover, the dots are blue or red whether they are inside or outside the safe voltage range, respectively.\\
Scenario \textit{a} shows some voltage problems: some over-voltages on the left-hand side of the plot, corresponding to more sunny days, and some under-voltages on the right-hand side, corresponding to the winter period.\\
It is possible to see that in all the control scenarios the agents are able to solve the voltage problems. In particular, in Scenario \textit{b} the voltage distribution of the nodes is close to 1. This is obtained only thanks to an excessive curtailment of the customers. In Scenarios \textit{c} and \textit{d} the agents solve the voltage problems while keeping the curtailment not too high. In both scenarios, the maximum and minimum voltage are close to the boundaries but inside the safe range, making the agents choosing actions without overdoing it. 
% It is also interesting to note that these values are low thanks to both active and reactive power control of the SIs. In case of only active power curtailment, the curtailed power would have been much higher and moreover the under-voltage problems would have not been solved.
It is worth noting that the low values of active power curtailment obtained in the different scenarios, can be attributed to the efficient control of both active and reactive power by the SIs. It should be emphasised that if only active power curtailment had been utilised, the amount of curtailed power would have been substantially greater, and the resolution of under-voltage issues would not have been achieved, as evident in other studies (\cite{IVRUDN, DVVCPVI, MFAFSPCURF}).

\begin{figure}[h!]
    \centering 
    \includegraphics[width=0.48\textwidth]{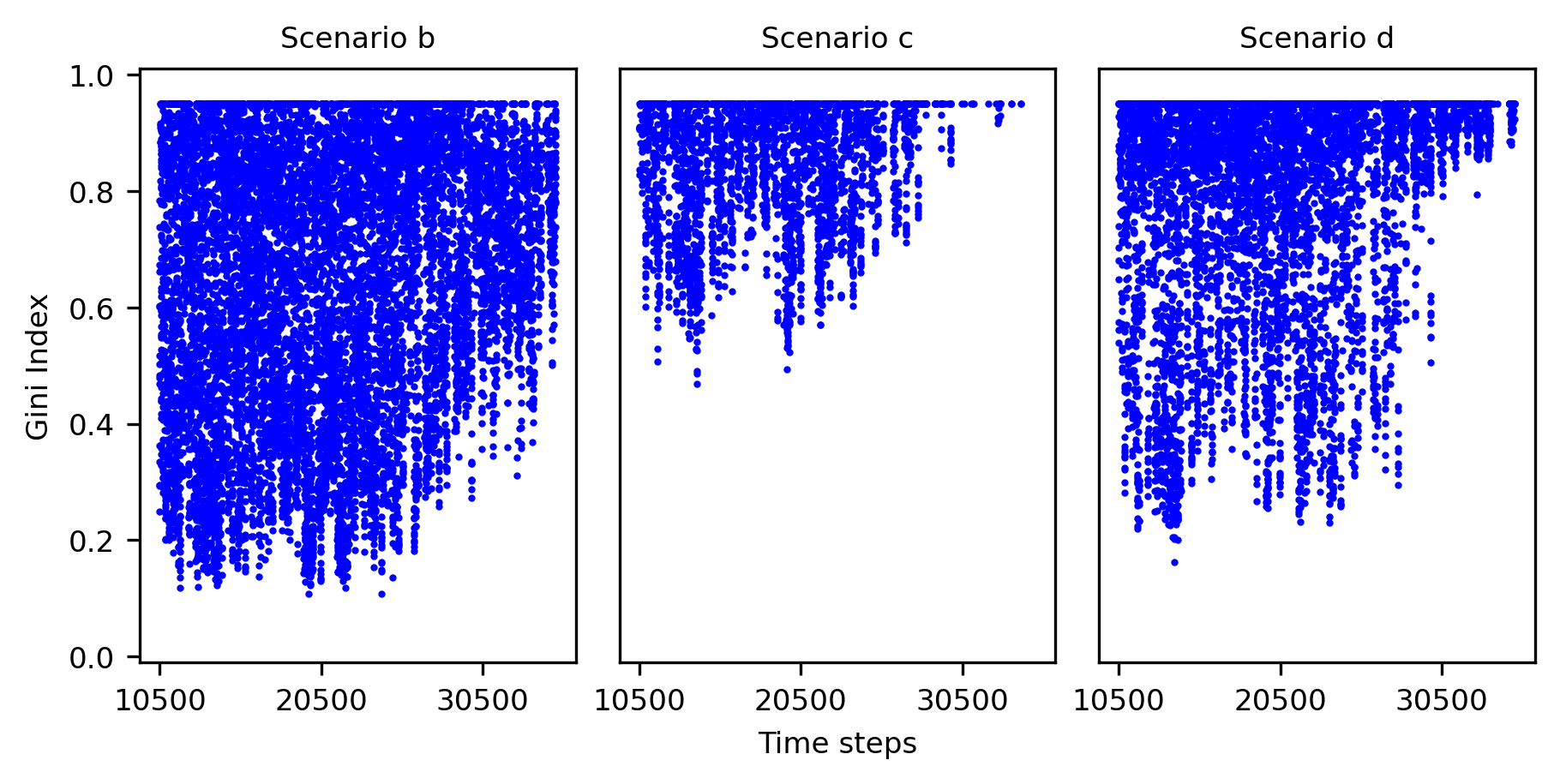}
    \caption{Gini index for each scenario where some control is applied}
    \label{fig:GI}
\end{figure}

The fairness of the curtailment in each scenario can be evaluated using the Gini index, as shown in Fig. \ref{fig:GI}. The figure shows the Gini index of the customers' curtailment for each time step where at least one customer is being curtailed. In Scenario \textit{b}, where only the voltage deviation punishment is considered, the curtailment appears to be fair, likely due to the large amount of curtailment applied. In this scenario, the total curtailment is 20.7 MWh equivalent to 44\% of the total production. In Scenario \textit{c}, the curtailment has a higher Gini index for most of the time steps, leading to a less fair result than in Scenario \textit{b}. In this scenario, the energy curtailed is 2.65 MWh equivalent to 5.6\% of the total production. Scenario \textit{c} required less curtailment to solve the voltage problems, but this means that often some customers are curtailed more than others. In Scenario \textit{d}, the curtailment is more fair than in Scenario \textit{c}, since the amount of power needed to be curtailed is distributed among all customers, even if this requires more curtailment overall. In this scenario, the energy curtailed is 8.77 MWh, equivalent to 18\% of the total energy produced.\\

It is important to note that, the agent in Scenario \textit{d} presents some limitations. In particular, from Fig. \ref{fig:GI}, it is possible to see that the curtailment is never totally fair, since the Gini index is never 0 (the minimum observed Gini index value is 0.16). At the same time, for some time steps the curtailment is unfair since the Gini index value is high (the maximum observed Gini index value is 0.95, which corresponds to only one customer being curtailed) and the agent behaves similarly to the other scenarios where fairness is not considered.

\begin{table}[h!]
\captionsetup{font=footnotesize}
\renewcommand{\arraystretch}{1.4}
\caption{\\Summary of the results for each scenario}
\label{tab:SS}
\centering
\resizebox{0.472\textwidth}{!}{%
\begin{tabular}{|c|c|c|c|c|c|c|}
\hline
\textbf{Scenario} &
  \textbf{\begin{tabular}[c]{@{}c@{}}Total PV\\ generation\\ (MWh)\end{tabular}} &
  \textbf{\begin{tabular}[c]{@{}c@{}}Energy\\ Curtailed\\ (MWh)\end{tabular}} &
  \textbf{\begin{tabular}[c]{@{}c@{}}\# Under\\ Voltages\end{tabular}} &
  \textbf{\begin{tabular}[c]{@{}c@{}}\# Over\\ Voltages\end{tabular}} &
  \textbf{\begin{tabular}[c]{@{}c@{}}Min V \\ (pu)\end{tabular}} &
  \textbf{\begin{tabular}[c]{@{}c@{}}Max V \\ (pu)\end{tabular}} \\ \hline
  
\textbf{\textit{a}} & \multirow{4}{*}{47.5} & 0    & 493 & 1677 & 0.9263 & 1.1191 \\ \cline{1-1} \cline{3-7} 

\textbf{\textit{b}} &                      & 20.7 & 0   & 0    & 0.9582 & 1.0321 \\ \cline{1-1} \cline{3-7} 

\textbf{\textit{c}} &                      & 2.65 & 0   & 0    & 0.9530 & 1.0497 \\ \cline{1-1} \cline{3-7} 

\textbf{\textit{d}} &                      & 8.77 & 0   & 0    & 0.9520 & 1.0409 \\ \hline
\end{tabular}%
}
\end{table}

Table \ref{tab:SS} shows a summary of the results obtained in each scenario. In particular, it shows the total PV energy produced by the customers in the testing period of approximately eight months, the energy curtailed by the RL agent, the number of time steps that present a voltage problem, either under or over-voltage, and the minimum and maximum voltage recorded at any bus in the eight months. It is possible to see that the agents in scenarios \textit{c} and \textit{d} solved all the voltage problems while keeping the energy curtailed low.\\

It is worth noting that the weights of the single terms have a great impact on the reward function and so on the results. In this experiment, the weights for the magnitude of the actions, $\beta$, and the fairness, $\omega$, are chosen to have similar impacts on the reward function as discussed in Section \ref{implementation}. However, in different situations, either $\beta$ or $\omega$ may have a greater impact depending on the specific goals of the DSO. Ultimately, it is a trade-off between how much active power curtailment the DSO is willing to accept without considering fairness versus increasing fairness at the cost of higher active power curtailment.

% \begin{figure}[h!]
% \centering
%     \subfloat[\label{fig:crit_sit_bef}]        {\includegraphics[width=0.9\linewidth]{figures/CritialSituationsBefore(1).png}}\\
    
%     \subfloat[\label{fig:crit_sit_aft}]       {\includegraphics[width=0.9\linewidth]{figures/CritialSituationsAfter(1).png}}\\

%     \caption[Agent's solution to voltage problems]{The buses' voltage situation before (a) and after (b) the agent's actions.}
%     \label{fig:Vsolv}
% \end{figure}

\section{Conclusion}\label{conclusion}
This paper investigates the control of SIs using a reinforcement learning technique to solve voltage problems in a power distribution network.
The model used is a deep deterministic policy gradient algorithm. This algorithm is able, using only the network’s information, to solve the critical situations in the network. In particular, the agent is able to solve the voltage problems, with a cost for the distribution system operator of a few MWh of energy not produced in the entire time span considered.\\
Moreover, the focus of this paper is solving the voltage issues with a fair curtailment strategy. The results show that it is possible to obtain a fair active power curtailment for the customers, controlling the active and reactive power of PV smart inverters, but at the cost of a higher curtailment.\\ 

Some future works may examine more complex networks; studying the impact of the weights $\beta$ and $\omega$ on the results; using a decentralised reinforcement learning technique to solve the voltage problems in a fair way, and considering the limitations of SIs.

\section*{References}
\printbibliography[heading=none]

\end{document}